\documentstyle[preprint,aps]{revtex}

\newcommand{\gsi}        {$\rm^{1}$}
\newcommand{\clt}        {$\rm^{2}$}
\newcommand{\heidelberg} {$\rm^{3}$}
\newcommand{\warsaw}     {$\rm^{4}$}
\newcommand{\bucarest}   {$\rm^{5}$}
\newcommand{\zagreb}     {$\rm^{6}$}
\newcommand{\itep}       {$\rm^{7}$}
\newcommand{\budapest}   {$\rm^{8}$}
\newcommand{\korea}      {$\rm^{9}$}
\newcommand{\dresde}     {$\rm^{10}$}
\newcommand{\kur}        {$\rm^{11}$}
\newcommand{\ires}       {$\rm^{12}$}

\begin{document}
\tightenlines


\title{Sideward flow of 
${\rm K}^+$ mesons in Ru+Ru  and Ni+Ni reactions \\
near threshold}

\author{P.~Crochet\gsi$^{\rm ,}$\clt\thanks{at LPC Clermont-Ferrand since 
01/10/98},
N.~Herrmann\gsi$^{\rm ,}$\heidelberg,
K.~Wi\'{s}niewski\gsi$^{\rm ,}$\warsaw,
Y.~Leifels\gsi$^{\rm ,}$\heidelberg,
A.\,Andronic\gsi,
R.~Averbeck\gsi,
A.~Devismes\gsi,
C.~Finck\gsi,
A.~Gobbi\gsi,
O.~Hartmann\gsi,
K.D.~Hildenbrand\gsi, 
P.~Koczon\gsi,
T.~Kress\gsi,
R.~Kutsche\gsi,
W.~Reisdorf\gsi,
D.~Sch\"ull\gsi, 
J.P.\,Alard\clt,
V.\,Barret\clt,
Z.\,Basrak\zagreb,
N.\,Bastid\clt,
I.\,Belyaev\itep,
A.\,Bendarag\clt,
G.\,Berek\budapest,
R.\,\v{C}aplar\zagreb,
N.\,Cindro\zagreb,
P.\,Dupieux\clt,
M.\,D\v{z}elalija\zagreb,
M.\,Eskef\heidelberg,
Z.\,Fodor\budapest,
%
Y.\,Grishkin\itep,
B.\,Hong\korea,
J.\,Kecskemeti\budapest,
Y.J.\,Kim\korea,
M.\,Kirejczyk\warsaw,
M.\,Korolija\zagreb,
R.\,Kotte\dresde,
M.\,Kowalczyk\warsaw,
A.\,Lebedev\itep,
K.S.\,Lee\korea,              
V.\,Manko\kur,
H.\,Merlitz\heidelberg,
S.\,Mohren\heidelberg,
D.\,Moisa\bucarest,
W.\,Neubert\dresde,
A.\,Nianine\kur,
D.\,Pelte\heidelberg,
M.\,Petrovici\bucarest,
C.\,Plettner\dresde,
F.\,Rami\ires,
B.\,de Schauenburg\ires,
Z.\,Seres\budapest,
B.\,Sikora\warsaw,
K.S.\,Sim\korea,
V.\,Simion\bucarest,
K.\,Siwek-Wilczy\'nska\warsaw,
V.\,Smolyankin\itep,
A.\,Somov\itep,
M.\,Stockmeier\heidelberg,
G.\,Stoicea\bucarest,
M.\,Vasiliev\kur,
P.\,Wagner\ires,
D.\,Wohlfarth\dresde,
J.T.\,Yang\korea,
I.\,Yushmanov\kur,
A.\,Zhilin\itep\\
the FOPI Collaboration 
}

\address{
\gsi~Gesellschaft f\"ur Schwerionenforschung, Darmstadt, Germany\\
\clt~Laboratoire de Physique Corpusculaire, IN2P3/CNRS
and Universit\'{e} Blaise Pascal, Clermont-Ferrand, France\\
\heidelberg~Physikalisches Institut der Universit\"at Heidelberg, Heidelberg, Germany\\
\warsaw~Institute of Experimental Physics, Warsaw University, Poland\\
\bucarest~National Institute for Nuclear Physics and Engineering, Bucharest, Romania\\
\zagreb~Rudjer Boskovic Institute, Zagreb, Croatia\\ 
\itep~Institute for Theoretical and Experimental Physics, Moscow, Russia\\
\budapest~KFKI Research Institute for Particle and Nuclear Physics, 
Budapest, Hungary\\
\korea~Korea University, Seoul, South Korea\\
\dresde~Forschungszentrum Rossendorf, Dresden, Germany\\
\kur~Kurchatov Institute, Moscow, Russia \\
\ires~Institut de Recherches Subatomiques, IN2P3-CNRS and Universit\'e
Louis Pasteur, Strasbourg, France \\
}

\maketitle

\begin{abstract}
Experimental data on ${\rm K}^+$ meson and proton sideward 
flow measured with the FOPI detector at SIS/GSI in the reactions 
${\rm Ru+Ru}$ at 1.69$A$~GeV and ${\rm Ni+Ni}$ at 1.93$A$~GeV are presented. 
The ${\rm K}^+$ sideward flow is found to be anti-correlated (correlated)
with the one of protons at low (high) transverse momenta.
When compared to the predictions of a transport model, 
the data favour the existence of an in-medium repulsive 
${\rm K}^+$-nucleon potential.
\end{abstract}

\vspace{1.cm}


The possible modification of hadron masses and widths 
in hot and dense matter is a subject of considerable current interest.
In particular a large theoretical effort has been devoted 
to the investigations of the in-medium properties of kaons as they are 
important for understanding both chiral symmetry restoration 
and neutron star properties~\cite{kolirev}.
These studies, carried out using different approaches~\cite{approach}, 
converge qualitatively towards the common feature that in the nuclear medium 
the ${\rm K}^+$ mesons feel a week repulsive potential whereas the 
${\rm K}^-$ mesons feel a strong attractive potential.
Both potentials can be parametrized by a linear dependence on 
the nuclear density~$\rho_0$ up to 2 times~$\rho_0$.
Experimentally, the in-medium kaon properties at low densities can be
studied by the analysis of kaon-nucleus scattering~\cite{sca} and kaonic
atoms~\cite{kaonic} data.
The properties of kaons in a high density medium can only be 
investigated by means of heavy ion collisions. 
This is particularly relevant for beam energies of 1-2$A$~GeV
for which, according to transport model calculations~\cite{friman},
the central region of the collision reaches nuclear densities of 2-3~$\rho_0$ 
and stays in this high density phase for a relatively long time 
compared to higher beam energies.
This beam energy range is also best suited to study the kaon in-medium 
properties since it corresponds to kaon production below threshold or 
close-to-threshold~\cite{kolirev}.
As the kaon in-medium potential results in a slightly increased
${\rm K}^+$ mass and a strongly reduced ${\rm K}^-$ mass, one expects 
to observe an enhanced ${\rm K}^-$ yield (its production being energetically 
much easier) and a reduced ${\rm K}^+$ yield (its production being 
energetically more difficult).
The large ${\rm K}^-$ production cross-section observed
by the KaoS collaboration in Ni+Ni collisions has been interpreted 
as an evidence for a reduced ${\rm K}^-$ effective mass in the nuclear 
medium~\cite{barth97}.
On the other hand the kaon potential should repel ${\rm K}^+$ 
from nucleons and attract ${\rm K}^-$ towards nucleons.
This would influence the phase space populations by a widening 
(narrowing) of the ${\rm K}^+$ (${\rm K}^-$) transverse momentum and 
rapidity distributions.
First signs of these effects have been observed very recently by the KaoS and 
the FOPI collaborations~\cite{laue99,kw}. 
Finally, the collective flow of kaons, both the in-plane component 
(the scope of this paper) and the out-of-plane component~\cite{shin98},
is also recognized as relevant observables to probe the kaon 
potential and thus provides useful complementary information~\cite{liko_all}.
Substantial uncertainties about the strength of the in-medium potential,
especially its momentum dependence~\cite{fuchs_2,schaff00}, and about the
in-medium cross-sections~\cite{schaff00} motivate further theoretical 
investigations as well as more detailed data.

The first experimental data on ${\rm K}^+$ sideward flow
have been obtained by the FOPI collaboration in Ni+Ni reactions 
at 1.93$A$~GeV~\cite{jim_all}.
The data show a vanishing ${\rm K}^+$ flow in the representation 
of the mean in-plane transverse momentum {\it versus} rapidity.
The sensitivity of such data to in-medium effects is under intense debate.
According to~\cite{li98} the data clearly support the existence of a 
repulsive ${\rm K}^+$-nucleon mean field.
According to~\cite{branpa97,wang97} the sensitivity of the
observable to in-medium
effects is found to be less pronounced but a slightly repulsive potential
cannot be excluded from the comparison.
On the other hand, the sensitivity of ${\rm K}^+$ sideward flow to in-medium
effects was found in~\cite{fuchs_2} to be washed-out when a particular momentum
dependence of the potential is included in the calculations.
It was recently pointed-out in~\cite{david98} that
the lifetime of nuclear resonances used in the models might be partially
responsible for the magnitude of the ${\rm K}^+$ sideward flow as it 
strongly affects the kaon production characteristics.

In order to further elucidate these questions, we investigate in this paper
the transverse momentum dependence of ${\rm K}^+$ and proton 
sideward flow in Ru+Ru and Ni+Ni systems.
Such a transverse momentum differential analysis does reveal more information 
than the transverse momentum integrated data where part of 
the effects are hidden.
In addition, a heavier system than Ni+Ni is better suited for flow studies 
since flow effects
are found to be larger, at least for baryons, as compared to lighter 
systems~\cite{willy}.
It allows also to study ${\rm K}^+$ flow in non-central collisions
where, due to a large sensitivity of the observable to in-medium effects,
an {\it anti-flow} phenomenon (see later) is expected to be 
seen~\cite{li98,casrep}.

The FOPI detector~\cite{gob93,jimnpb} is an azimuthally symmetric apparatus
made of several sub-detectors which provide charge and mass determination
over nearly the full 4$\pi$ solid angle.
For the analysis presented here, only the Central Drift Chamber (CDC),
the time of flight array (Barrel) and the forward Plastic Wall (PLA) were
used.
The CDC and the Barrel are placed in a solenoidal magnetic field of
0.6~${\rm T}$.
Pions, kaons, protons, deuterons and tritons are identified
with the CDC ($33^{\circ}<\theta_{\rm lab}<150^{\circ}$) 
by measurement of the specific energy loss in the CDC gas and the 
magnetic rigidity.
Due to the low kaon yield in this beam energy regime, additional 
redundancy for kaon identification is needed.
This is achieved by adding to the previous informations
the particle velocity which is determined from the extrapolation
of a track in the CDC to the appropriate hit in the Barrel.
The acceptance is therefore reduced for kaons to the geometrical coverage 
of the Barrel~: $39^{\circ}<\theta_{\rm lab}<130^{\circ}$.
Kaon detection is possible only for transverse momenta above
$p_t=0.1~{\rm GeV/c}$ which is needed for a particle to reach the Barrel.
The upper momentum limit to which ${\rm K}^+$ can be identified without 
significant contamination from pions and protons is 
$p_{\rm lab}=0.5~{\rm GeV/c}$.
More details about kaon identification with the FOPI detector can be found 
in~\cite{kw,jim_all,jimnpb,best}.

\begin{figure}[hhh]
\includegraphics{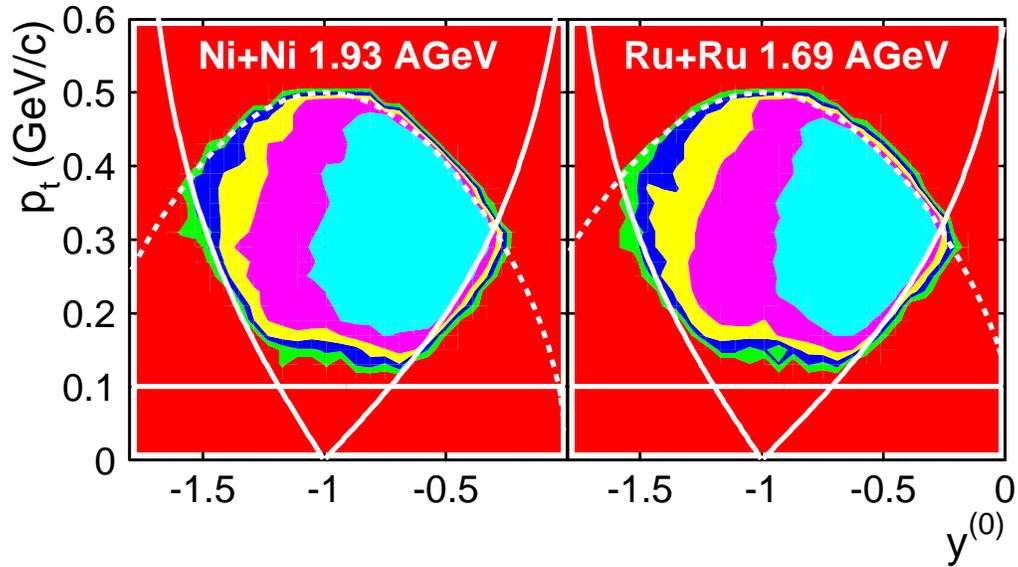}
\vspace*{15.cm}
\caption{ Measured yield of ${\rm K}^+$ in a plane
transverse momentum $p_t$ {\it versus} normalized
rapidity $y^{(0)}$ in the reactions Ni+Ni at 1.93$A$~GeV (left) and Ru+Ru 
at 1.69$A$~GeV (right).
The contour levels correspond to logarithmically increasing intensity. 
The solid curves denote the geometrical limits of the detector 
acceptance ($\theta_{\rm lab}=39^{\circ}$ and $130^{\circ}$).
The dashed curve corresponds to $p_{\rm lab}=0.5~{\rm GeV/c}$.
The solid horizontal line corresponds to $p_t=0.1~{\rm GeV/c}$.
Here, the yield is neither corrected for kaon decay losses nor for detection 
efficiency.}
\label{pty}
\end{figure}

The acceptance of the FOPI detector for ${\rm K}^+$ identification 
is shown in Fig.~\ref{pty} in 
terms of ${\rm K}^+$ transverse momentum as a function of the normalized
rapidity $y^{(0)}$ where $y^{(0)}$ denotes the particle rapidity divided 
by the beam rapidity in the center-of-mass (c.m.) system.
With this normalization, -1,0 and 1 correspond to target, c.m. and projectile
rapidity, respectively.

The events were centrality selected by imposing conditions on the 
multiplicity
PMUL~\cite{ala92} of charged particles detected in the outer part of
the Plastic Wall ($7^{\circ}<\theta_{\rm lab}<30^{\circ}$).
For the Ni+Ni system, one class of central events was selected
whereas for the Ru+Ru system, a central and a semi-central event classes 
were considered.
The features of these event classes are listed in 
Table~\ref{tab}.
The reaction plane was reconstructed event-wise, according to
the method devised in~\cite{dan85}.
In order to remove autocorrelation effects, the azimuth of the reaction
plane was estimated for each particle in a given event
using all detected baryons in the event except the particle of interest.
The flow observable presented here was corrected for the accuracy 
with which the reaction plane was determined,
according to the method described in~\cite{oll97}.
The corresponding correction factors $f$ are shown in Table~\ref{tab}.

The $p_t$ dependence of the sideward flow has been investigated 
by means of a Fourier expansion of azimuthal distributions.
$\phi$ being the azimuthal angle of a particle with respect to the reaction
plane, the azimuthal distributions $dN/d\phi$ can be parametrized by 
$\sim (1 + 2 v_1\cos(\phi) + 2 v_2\cos(2 \phi) +...)$,
where $v_n=<\cos(n\phi )>$ are the Fourier coefficients.
Sideward flow is related to the first Fourier coefficient by~:
$v_1=<\cos(\phi)>=<p_x/p_t>$ where $p_x$
is the transverse momentum projected onto the reaction plane
(for more details, see~\cite{olqm97,vol}).
In order to exclude of non-trivial distortions introduced by the detector
acceptance for ${\rm K}^+$, $v_1$ was extracted in a portion 
of phase space free of any geometrical bias. 
From Fig.~\ref{pty} it can be seen that requiring $-1.2<y^{(0)}<-0.65$
defines the $p_t$ window $0.15 < p_t < 0.45$ where 
$v_1$ can be extracted without any geometrical acceptance effects for both 
systems.
Note that the $p_t$ range extends to much higher values for protons.

The magnitude of the ${\rm K}^+$ sideward flow signal was found to depend 
somewhat on the mass window used to select ${\rm K}^+$
candidates and the applied quality criteria for the tracks in the 
CDC and their matching with the Barrel detector.
Therefore, systematical uncertainties were estimated by adding quadratically
the errors estimated by comparing the flow values obtained with 
``strong" and ``open" selection criteria.
The corresponding boundaries of these selection criteria were 
established from the apparatus resolution on the one hand, and from the 
degree of contamination of ${\rm K}^+$ by other particles on the other hand.
These systematical errors are smaller than the size of the symbols used
in the following figures.
Other possible sources of systematical errors have been investigated for the 
Ru+Ru system by means of Monte-Carlo simulations using the
GEANT package~\cite{rbrun}.
This consists of a complete simulation of the FOPI apparatus 
including resolutions in energy deposition and spatial position, front-end 
electronic processing, hit reconstruction, hit tracking and track 
matching between the sub-detectors. 
The output of GEANT was analyzed in the same way as the experimental 
data and then compared to the input of the simulation.
The results of the full simulation overestimate $v_1$ by few percent  
in the region of the phase space under consideration.
This systematical effect was found to be {\it i)} independent
of $p_t$, {\it ii)} slightly more pronounced in semi-central collisions
than in central collisions and {\it iii)} almost negligible for protons.
It is attributed to a loss of particles in the azimuthal 
region of the high track density of the reactions.
Based on these simulations, the data points in 
Fig.~\ref{flow_cent_ru1690_new}
have been shifted-down by the offset $S$ reported in Table.~\ref{tab}.
Since no signs of such systematical bias have been observed in the Ni+Ni
system, for which the track density is significantly lower, no correction 
has been applied to the data for this system.
The ${\rm K}^+$ data points in the two following figures have 
been corrected for the kaon decay losses although this effect was found to be 
negligible on $v_1$ in the region of phase space considered.

\begin{figure}[hhh]
\includegraphics{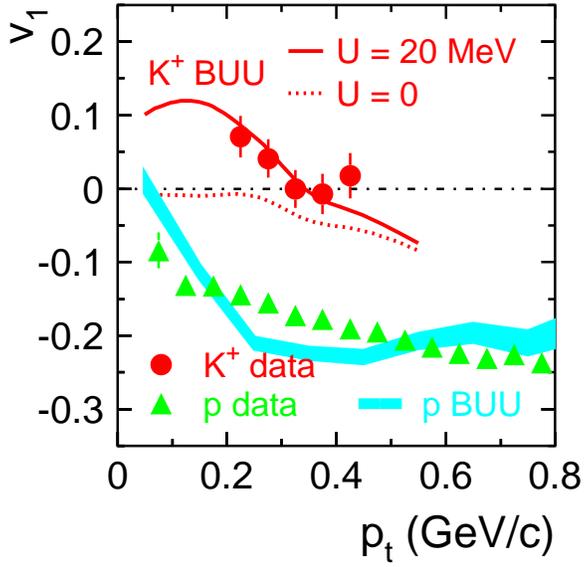}
\vspace*{15.cm}
\caption{$v_1$ {\it versus} $p_t$ for protons (triangles) 
and ${\rm K}^+$ (dots) in the rapidity range -1.2$<y^{(0)}<$-0.65
for central Ni+Ni reactions at 1.93$A$~GeV.
Error bars represent statistical uncertainties.
The curves and shaded area show the predictions of the RBUU model 
for ${\rm K}^+$ and proton, respectively.
The statistical uncertainties on RBUU-${\rm K}^+$ flow are similar to the ones 
on RBUU-proton flow. The latter are represented by the width of the
shaded area.}
\label{v1_ni_new}
\end{figure}

\begin{figure}[hhh]
\includegraphics{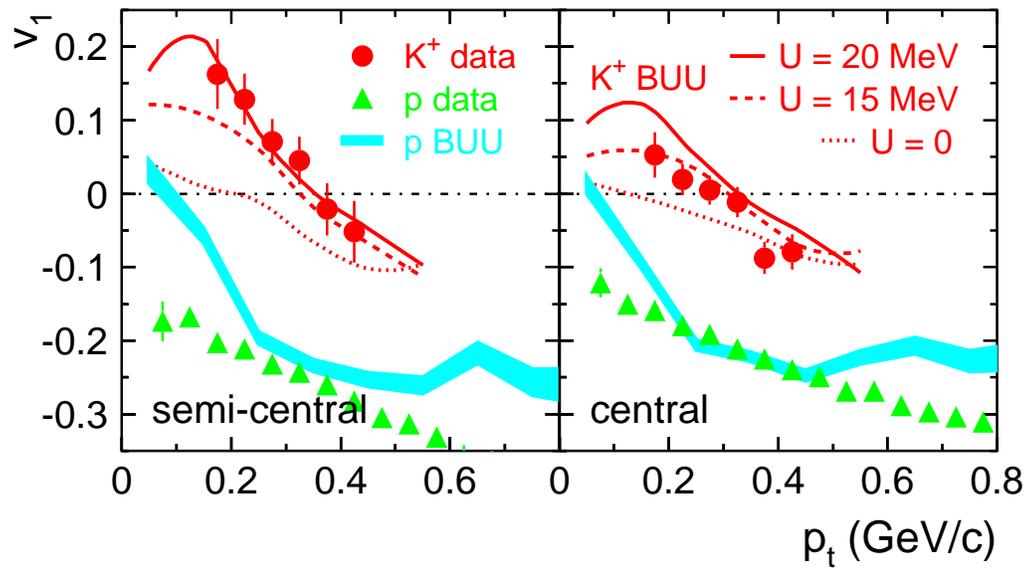}
\vspace*{15.cm}
\caption{Same as Fig.~\ref{v1_ni_new} for 
semi-central (left) and central (right)
Ru+Ru reactions at 1.69$A$~GeV.}
\label{flow_cent_ru1690_new}
\end{figure}

The ${\rm K}^+$ and proton sideward flow is shown in
Fig.~\ref{v1_ni_new} for the Ni+Ni system.
It can be seen that the ${\rm K}^+$ flow pattern is 
totally different than the one of protons. 
Protons have a negative $v_1$ for all $p_t$.
Since the rapidity window used is located in the backward hemisphere, 
this means that protons are positively flowing whatever their $p_t$.
In contrast, ${\rm K}^+$ have positive $v_1$ for low $p_t$.
In other words, ${\rm K}^+$ are negatively flowing (or anti-flowing) at
low $p_t$  while their $v_1$ is compatible with 0 at large $p_t$.
We stress that vanishing ${\rm K}^+$ flow was seen, if 
$p_t$-integrated data were used~\cite{jim_all}.
This demonstrates the need to study flow effects simultaneously 
in a $p_t$-integrated and $p_t$-differential way.
Investigating the flow differentially has in addition the advantage that 
comparison of data to model predictions is straightforward since no 
corrections are necessary for the finite acceptance of the apparatus.

The main features of the ${\rm K}^+$ flow pattern are exhibited more
clearly for the heavier system Ru+Ru for which the centrality dependence 
of $v_1$ is shown in Fig.~\ref{flow_cent_ru1690_new}.
Here there is a clear trend for positive $v_1$ (anti-flow) at low $p_t$
and negative $v_1$ (flow) at large $p_t$.
A change in the ${\rm K}^+$ flow pattern can 
be observed, from central to semi-central collisions.
Note also the change in the proton flow pattern and in the
difference between the ${\rm K}^+$ and the proton signals.
It has been shown in~\cite{crogsi}, that in data averaged over $p_t$
no ${\rm K}^+$ flow is seen in central Ru+Ru reactions, while some 
antiflow is observed in semi-central events.
A similar anti-flow pattern has been observed very recently 
for ${\rm K}^0_{\rm S}$
in Au+Au collisions at 6$A$~GeV~\cite{rai99}.

In order to demonstrate the sensitivity of the experimental 
findings to the properties of kaons in dense hadronic matter,
the data were compared to the predictions of two different realisations 
of the Relativistic Boltzmann-Uehling-Uhlenbeck (RBUU) model~\cite{casrep}~: 
without and with in-medium effects.
The first situation corresponds to a calculation including 
binary collisions plus potentials except kaon potentials.
In the second scenario in-medium effects are taken into account.
They are introduced by means of a dispersion relation
from which kaon effective potentials and masses are derived.
This results, for ${\rm K}^+$, in an increased effective mass and 
a repulsive potential.
The former tends to lower the ${\rm K}^+$ production probability in a first 
chance nucleon-nucleon collision while the latter tends to push
${\rm K}^+$ away from nucleons.
The strength of the in-medium ${\rm K}^+$ potential at normal nuclear density
was fixed to ${\rm U}=15~{\rm MeV}$ and $20~{\rm MeV}$
for the Ru+Ru system and to ${\rm U}=20~{\rm MeV}$ for the 
Ni+Ni system.
More details about the calculations can be found in~\cite{branpa97,casrep}.
The centrality selection criteria imposed on the data described above was 
modeled by an impact parameter selection of the RBUU events requesting the 
same geometrical cross section.
The data points from RBUU were extracted in the same rapidity window 
as the one used for the experimental data.
No further conditions were applied to the calculations since {\it i)}
the transverse momentum range $0.15 < p_t < 0.45$ defined with the previously 
discussed rapidity window is free of any detector bias and {\it ii)} the 
experimental data points are corrected for the reaction plane fluctuations, 
the detection inefficiency and the kaon decay losses.

The results of the calculations are shown by the curves in 
Fig.~\ref{v1_ni_new} and~\ref{flow_cent_ru1690_new}.
It can be observed that without in-medium ${\rm K}^+$ potential the 
calculation fails to describe the low-$p_t$ ${\rm K}^+$ 
anti-flow phenomenon observed in the data.
In contrast, when in-medium effects are taken into account the model 
reproduces quantitatively ${\rm K}^+$ experimental signals for both 
systems.
The additional repulsive potential pushes ${\rm K}^+$ further away from 
nucleons therefore resulting in an anti-correlation between the ${\rm K}^+$
flow and the proton flow.
It is important to mention that neither rescattering effects nor 
the Coulomb repulsion can explain satisfactorily the experimental behaviour
of ${\rm K}^+$ flow, since both of them are included 
independently of the in-medium potential.
Furthermore, rescattering of ${\rm K}^+$ with nucleons is expected to 
increase slightly the ${\rm K}^+$ sideward flow in the direction of 
nucleons~\cite{li98}, and Coulomb potential is found to play an almost 
negligible role on ${\rm K}^+$ sideward flow~\cite{wang97}.

The results obtained here are in good qualitative agreement with the 
predictions of another independent transport model calculation including 
similar in-medium effects~\cite{li98}.
In addition, it has been shown that these two calculations give a consistent 
description of the measured ${\rm K}^-/{\rm K}^+$ ratio, for the same
reactions, only if in-medium effects are taken into 
account~\cite{kw,casrep,li982}.

On the other hand, the model fails in consistently describing the proton 
sideward-flow data in the considered target rapidity region, although a 
reasonable agreement is found in the mid-rapidity region~\cite{crogsi}.
This discrepancy is mostly due to an improper separation of free protons 
and bound nucleons in the target spectator~\cite{caspriv} which seems to 
be a general problem of transport model calculations.
A similar discrepancy has indeed been observed from the comparison of 
experimental data and the predictions of the Relativistic Quantum Molecular 
Dynamics model in Au+Au reactions at 11$A$~GeV~\cite{e887}.
This shows that more definite interpretation of the $p_t$-differential flow 
data for nucleon needs further detailed investigations.
                                            
Due to the discrepancies of the $p_t$-dependence of the baryon flow,
at this moment, no final conclusion about the strength of the kaon 
potential can be drawn although, model-dependently, the errors on the data
points would allow for an accuracy of about 
10~MeV (see Fig.~\ref{flow_cent_ru1690_new} left).
There is, however, no other mechanism but a repulsive potential that
would allow to generate the observed antiflow of ${\rm K}^+$ at low $p_t$.

In summary the transverse momentum and centrality dependence of ${\rm K}^+$
and proton sideward flow in Ni+Ni and Ru+Ru collisions at SIS energies have 
been studied with the FOPI detector.
The data near target rapidity reveal a ${\rm K}^+$ anti-flow phenomenon 
originating mostly from low $p_t$ ${\rm K}^+$.
The comparison of the data with the predictions of a transport model
investigating in-medium kaon properties clearly favour the existence
of an in-medium repulsive potential for ${\rm K}^+$.
The study of ${\rm K}^-$ flow, for which in-medium effects are expected 
to be more pronounced, should shed more light on this issue.

\acknowledgements

We gratefully acknowledge E.L.~Bratkovskaya and W.~Cassing for providing 
us the RBUU calculations and for fruitful discussions. 
This work was supported by the agreement between
GSI and IN2P3/CEA and by the PROCOPE-Program of the DAAD. 
The BMBF supplied support under the contracts RUM-005-95, POL-119-95,
UNG-021-96 and RUS-676-98,
the DFG within the projects 436~RUS-113/143/2 and 446~KOR-113/76/0.
Support has been received from the Polish KBN under grant 2P302-011-04, 
from the Korean KOSEF under grant 985-0200-004-2, 
from the Hungarian OMFB under contract D-86/96 and from the 
Hungarian OTKA under grant T029379.

\begin{center}
\begin{table}[h]
\caption{Number of recorded events (${\rm N}_{evt}$),
number of identified ${\rm K}^+$ (${\rm N}_{{\rm K}^+}$),
mean geometrical impact parameter ($b_{\rm geom}$), 
correction factor ($f$) and offset ($S$) to sideward flow observables for
the selected classes of events (see text). 
$b_{\rm geom}$ was calculated assuming a sharp cut-off 
approximation~\protect\cite{cav90}.}
\begin{tabular}{llll}
System     & Ni+Ni  & Ru+Ru  & Ru+Ru \\ 
Centrality     &  central  &  semi-central & central \\ 
\hline
${\rm N}_{\rm evt}$ $(\times 10^6)$    & 1.4     & 1.8  & 4.5 \\
${\rm N}_{{\rm K}^+}$                  & 12700   & 5000 & 15900 \\
$b_{\rm geom}$ (fm)                    &  1.7    & 3.8  & 2.3 \\
$f$                                    &  1.48   & 1.18 & 1.28 \\
$S_{\rm proton}$                       &  0.     & 0.02 & 0.01 \\
$S_{{\rm K}^+}$                        &  0.     & 0.04 & 0.03 \\
\end{tabular}
\label{tab}
\end{table}
\end{center}

\newpage

\newpage

\newpage

\end{document}